\documentstyle[11pt,paspconf,epsf]{article}

\begin{document}

\title{Global textures and the Doppler peaks}
\author{Alejandro Gangui\altaffilmark{1}}
\affil{{\sl ICTP} -- International Center for Theoretical Physics, 
Trieste, Italy}
\author{Ruth Durrer and Mairi Sakellariadou}
\affil{D\'ept de Physique Th\'eorique, Universit\'e de Gen\`eve, Switzerland}

\altaffiltext{1}{Also at {\sl SISSA} -- International School for Advanced 
Studies, gangui@gandalf.sissa.it}

\begin{abstract}
We review recent work aimed at showing how
global topological defects influence the shape of the 
angular power spectrum of the cosmic microwave background 
radiation on small scales.
While Sachs--Wolfe fluctuations give the dominant contribution 
on angular scales larger than about a few degrees, 
on intermediate scales,
$0.1^\circ
\mathrel{\raise.3ex\hbox{$<$\kern-.75em\lower1ex\hbox{$\sim$}}}
\theta
\mathrel{\raise.3ex\hbox{$<$\kern-.75em\lower1ex\hbox{$\sim$}}}
2^\circ$,
the main r\^ole is played by coherent oscillations in
the baryon radiation plasma before recombination.
In standard cosmological models
these oscillations lead to the  `Doppler peaks' in the angular
power spectrum.
Inflation--based cold dark matter models predict the location of 
the first peak to be 
at $\ell \sim 220 / \sqrt{\Omega_0}$, with a height which is a few times the
level of anisotropies at large scales.
Here we focus on perturbations induced by global textures.
We find  that  the height of the first peak is reduced
and is shifted to $\ell\sim 350$.
\end{abstract}


\section{Introduction}

The importance of the observations of anisotropies on intermediate 
and small angular scales cannot be over-emphasized.
Presently a number of experiments scrutinize different regions of the 
sky trying to uncover the real amplitude and structure of the relic radiation. 
%
The CMB anisotropies are a powerful tool to discriminate among
the different models of structure formation  by  purely linear analysis.
These anisotropies are parameterized in terms of
$C_\ell$'s, defined 
as the coefficients in the expansion of the angular 
correlation function
\begin{equation}
\label{twotwo}
\langle
C_2(\vartheta)
\rangle
= {1\over 4\pi}\sum_\ell(2\ell+1)C_\ell P_\ell(\cos\vartheta). 
\end{equation}
For Harrison--Zel'dovich spectra of perturbations $\ell (\ell + 1) C_\ell$
is constant on large angular scales, say 
$\ell 
\mathrel{\raise.3ex\hbox{$<$\kern-.75em\lower1ex\hbox{$\sim$}}}
50$.
Both inflation and topological defect models lead to approximately scale
invariant spectra on large scales.
The next important step towards discriminating between competing models is
the measurement of degree scale anisotropies.  
Disregarding Silk damping, gauge invariant linear perturbation 
analysis leads to (Durrer 1994)
\begin{equation} 
{\delta T\over T} = \left[
 - {1\over 4}D_g^{(r)} + V_j \gamma^j -\Psi+\Phi\right]_i^f + 
\int_i^f (\Psi' - \Phi' ) d\tau  ~, 
\label{dT} 
\end{equation}
where $\Phi$ and $\Psi$ are quantities describing the perturbations 
in the geometry and $\bf V$ denotes the peculiar velocity of 
the radiation fluid with respect to the overall Friedmann expansion.
In Eq.(\ref{dT}), $D_g^{(r)} = \delta\rho_r/\rho_r$ 
specifies the intrinsic density fluctuation in the radiation fluid; 
below we use
$D_r = (\delta\rho_r +\delta\rho_b)/(\rho_r+\rho_b)$, 
about 5\% smaller than $D_g^{(r)}$ for scales inside the horizon,
which are the ones we are interested in.

The Sachs--Wolfe (SW) contribution in the above equation 
has been calculated for both inflation and defect models,
yielding mainly a normalization of the spectra.
We present below a computation for the Doppler 
contribution from global topological defects; 
in particular we perform our analysis for $\pi_3$--defects, 
textures (Turok 1989), in a universe dominated by 
cold dark matter (CDM). 
We believe that the main ideas of our analysis remain valid for 
all global defects.
The Doppler contribution to the CMB anisotropies is approximately 
given by\footnote{We are neglecting here the SW effect, 
which decays on subhorizon scales like $\ell^{-2}$. 
We also neglect the contribution of the neutrino fluctuations.
These approximations lead to an error of less than about 30\% in the 
amplitude of the first Doppler peak; 
see discussion in (Durrer, Gangui \& Sakellariadou 1995).}
\begin{equation}
\left[
{\delta T \over T}({\vec x},\hat\gamma )
\right]^{Doppler} \approx 
{1\over 4}D_r({\vec x}_{rec},\eta_{rec}) -
{\vec V}({\vec x}_{rec},\eta_{rec})\cdot {\hat\gamma} ~,
\label{Doppler}
\end{equation}
where ${\vec x}_{rec} = {\vec x} + {\hat\gamma} \eta_0$.  
In the previous formula ${\hat\gamma}$ denotes a direction in the sky
and $\eta $ is the conformal time, with $\eta_0 $ and $\eta_{rec} $ 
the present and recombination times, respectively.

\section{Linear theory power spectra}
\label{sec-linearspectra}

We now study a two--fluid system: 
baryons plus radiation, which prior to
recombination are tightly coupled, and CDM.
We start off by considering the evolution equations
%
in each fluid component 
as given by Kodama \& Sasaki (1984).
The evolution of the perturbation variables 
in a flat background, $\Omega = 1$,  can be expressed as 
\begin{equation} 
\begin{array}{lll}
V'_r +{a'\over a}V_r &=& k\Psi+ k{c_s^2\over 1+w}D_r \\
V'_c +{a'\over a}V_c &=& k\Psi  \\
D'_r - 3 w {a'\over a}D_r &=& (1+w) [ 3{a'\over a}\Psi-3\Phi' 
   -kV_r - {9\over2} \left({a'\over a}\right)^2 k^{-1}
   (1+{w\rho_r\over \rho})V_r  ] \\
D'_c                 &=& 3{a'\over a}\Psi-3\Phi' - k V_c
   -{9\over2} \left({a'\over a}\right)^2 k^{-1} 
   (1+{w\rho_r\over \rho})V_c
~, 
\end{array}
\label{KSeq} 
\end{equation}
where subscripts $_r$ and $_c$ denote the baryon--radiation plasma 
and CDM, respectively; 
$D,~V$ are density and velocity perturbations;
$w=p_r/\rho_r$, $c_s^2=p'_r/\rho'_r$ and $\rho = \rho_r + \rho_c$.

Throughout this section we will normalize the scale factor
such that  $a(\eta_{eq}) = 1$, where $\eta_{eq}$ is the 
conformal time at which matter and radiation densities coincide.
Setting $\tau \equiv (8\pi G \rho_{eq} / 6)^{-1/2}$,
where $\rho_{eq}$ is the total energy density at equality, 
we obtain 
$a(\eta) = (\eta / \tau)(1 + {1\over 4}(\eta / \tau))$.
It is clear from this expression that for early (late) conformal 
times we recover the radiation (matter) dominated phase of expansion
of the universe.

Since  the energy density of baryons and CDM 
go like $\propto a^{-3}$, the ratio  
$\rho_{B} / \rho_{c}$ is const and equal to 
$\Omega_{B}$. Therefore 
$(\rho_{B} / \rho_{rad}) \simeq \Omega_B a$.
This leads to 
\begin{equation} 
w     \simeq {1\over 3}(1+\Omega_B a)^{-1} ~~ ; ~~  
c_s^2 \simeq {1\over 3}(1+{3\over 4}\Omega_B a)^{-1}.
\end{equation}

The only place where the seeds enter the system (\ref{KSeq}) 
is through the potentials $\Psi$ and $\Phi$.
These potentials may be split into a part coming from standard  
matter and radiation, and a part due to the seeds, e.g.,  
$\Psi = \Psi_{(c,r)} + \Psi_{seed}$ where $\Psi_{seed},\Phi_{seed}$ 
are determined by the energy momentum tensor of the seed; the global
texture in our case.
Having said this, one  may easily see how the seed source terms 
arise.
Long but straightforward algebra leads to two second order
equations for $D_r$ and $D_c$, namely
\begin{eqnarray}
D_r''+{a'\over a}[1+3c_s^2-6w+F^{-1}\rho_c]D_r'
	 -{a'\over a}\rho_cF^{-1}(1+w)D_c' 
&&\nonumber \\
+4\pi Ga^2[\rho_r(3w^2-8w+6c_s^2-1)-2F^{-1}w\rho_c(\rho_r+\rho_c)
&&\nonumber\\
+\rho_c(9c_s^2-7w)+
{k^2\over 4\pi Ga^2}c_s^2]D_r -4\pi G a^2\rho_c(1+w)D_c 
&=&(1+w)S~;
  \nonumber \\
D_c''+{a'\over a}[1+(1+w)F^{-1}\rho_r(1+3c_s^2)]D_c'
	 -{a'\over a}(1+3c_s^2)F^{-1}\rho_rD_r' 
&&\nonumber \\
	-4 \pi G a^2\rho_cD_c -
	4 \pi G a^2\rho_r(1+3c_s^2)[1-2(\rho_r+\rho_c)F^{-1}w]D_r
&=&S~,  
\label{dc}
\end{eqnarray}
where $F\equiv k^2(12\pi Ga^2)^{-1}+\rho_r(1+w)+\rho_c$ and  
$S$ denotes a source term, 
which in general is given by $S=4\pi G a^2(\rho +3p)^{seed}$.
In our case, where the seed is described by a global scalar field $\phi$,
we have $S=8\pi G (\phi ')^2$.
From numerical simulations one finds that the average of $|\phi '|^2$
over a shell of radius $k$, can be modeled by (Durrer \& Zhou 1995)
\begin{equation}
\langle|\phi'|^2\rangle (k, \eta)\ =\  
{{1\over 2} A{\tilde \eta}^2\over \sqrt {\eta}[1+\alpha (k\eta)
+\beta (k\eta)^2]},  \label{fit}
\end{equation}
with ${\tilde \eta}$ 
denoting the symmetry breaking scale of the phase transition 
leading to texture formation.
The parameters in (\ref{fit}) are  
$A\sim 3.3$, $\alpha\sim -0.7/(2\pi)$ and $\beta \sim 0.7/(2\pi)^2$.
On super--horizon scales, where the source term is important,  
this fit is accurate to about $10\%$. 
On small scales the accuracy reduces to a factor of 2.  
By using this fit in the calculation of $D_r$ and $D_c$ from 
Eqs.(\ref{dc}) we effectively neglect the 
time evolution of phases of $(\phi ')^2$; 
the incoherent evolution of these phases may smear out subsequent 
Doppler peaks (Albrech et al. 1995), but will not affect substantially the
height of the first peak.

\section{Choosing initial conditions}
\label{sec-inicon}

In defect models of large scale structure formation
one usually assumes that the universe begins in a hot, 
homogeneous state.
All perturbation variables are zero, and the primordial 
fields (e.g., the global field $\phi$ in our case) 
are in thermal equilibrium in the unbroken symmetry phase.
When the phase transition occurs the original value $\phi =0$
becomes unstable and $\phi$ assumes a value which lies in the sphere 
of degenerate minima determined by $<\phi^2>=\tilde{\eta}^2$. 
Causality requires the correlation length to be bounded by the size
of the horizon.  On super--horizon scales the correlations vanish and 
therefore the defect stress energy tensor has a white noise power spectrum for
 $k \eta << 1$.
After the phase transition, the defect energy rapidly enters scaling,
$\rho_{def}\propto \eta^{-2}$.  
Adopting this white noise behaviour of the field
on super--horizon scales, one can simulate $\phi$ on scales much larger than 
the horizon scale at the phase transition.  The only condition is that the 
scales considered are super--horizon at the time one starts the simulation.
This is very important, since the scales relevant for large scale structure 
are more than 50 orders of magnitude larger than the inverse symmetry breaking
scale $\tilde{\eta}^{-1}$. 

Going back now to our set of 
Eqs.(\ref{dc}) we will specify our initial conditions
as follows: for a given scale $k$ we choose the initial time 
$\eta_{in}$ such that the perturbation is super--horizon and the 
universe is radiation dominated.
In this limit the evolution equations reduce to
\begin{equation}
D_r''-{2\over \eta^2} D_r ={4\over 3}{A\epsilon \over \sqrt {\eta}} 
~~~;~~~
D_c''+{3\over \eta} D_c'-{3\over 2\eta}D_r'
-{3\over 2\eta^2}D_r  = {A\epsilon \over \sqrt {\eta}} ~,
\end{equation}
with particular solutions
$D_r =  -{16\over 15}\epsilon A \eta^{3/2}$ and 
$D_c =  -{4 \over  7}\epsilon A \eta^{3/2}$.
In the above equations we have introduced 
$\epsilon\equiv 4\pi G{\tilde \eta}^2$, the only free parameter in the model. 
We consider perturbations  seeded by the texture field, and 
therefore it is incorrect to add a homogeneous growing mode to the 
above solutions.  
With these initial conditions, Eqs.(\ref{dc}) are easily 
integrated numerically, leading to the spectra for $D_r(k, \eta_{rec})$ 
and $D_r '(k, \eta_{rec})$. (see Durrer, Gangui \& Sakellariadou 1995 
for details).

\section{Angular power spectrum from global textures} 
\label{sec-textucls}

In this section we calculate the expression for $C_{\ell}$
as a function of the baryon--radiation power spectrum and its
derivative, evaluated at recombination,  $\eta_{rec}$.
We express the Fourier transform of the velocity perturbation
in the baryon--radiation component as 
${\vec V}({\vec k}) \simeq 
- i {\vec k} D_r'({\vec k}) / [ k^2 (1 + w) ]$.
We then Fourier transform the Doppler contribution to the
anisotropies given by Eq.(\ref{Doppler}) and perform a 
standard analysis to get the Doppler contribution to the 
angular power spectrum 
$C_{\ell}$. This is given as a function of 
the spectra for $D_r(k, \eta_{rec})$
and $D_r '(k, \eta_{rec})$ as follows
\begin{equation}
C_{\ell} = {2\over \pi} \int dk \left[{k^2\over 16}|D_r(k,t_{rec})|^2
j_{\ell}^2(kt_0) + (1+w)^{-2} |D_r'(k,t_{rec})|^2  (j_{\ell}'(kt_0))^2\right].
\label{cldopp}
\end{equation}
As we mentioned earlier, the source term in Eqs.(\ref{dc}) 
was estimated from numerical simulations 
of the evolution of the defect field, which 
give a fit for the absolute value of $\phi$, but do not specify the phases.
The crossed term missing in Eq.(\ref{cldopp}), cannot be determined with
this method.  However, due to phase averaging, we expect it to be diminished
substantially and therefore we neglect it. 
Integrating Eq.(\ref{cldopp}), we obtain the Doppler contribution to the
CMB anisotropies shown in Figure 1.

\begin{figure}
\plotfiddle{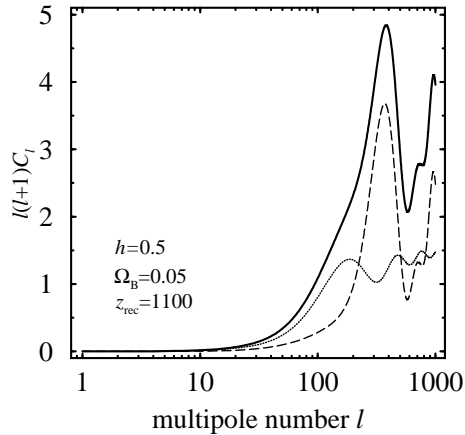}{6cm}{0}{35}{35}{-122}{-65}
\caption{The angular power spectrum for the Doppler contribution to
the  CMB anisotropies is shown in units of $\epsilon^2$ (thick solid curve).
Contributions from the baryon--radiation density perturbation (dashed curve)
and its derivative (dotted curve) from Eq.(9) are also shown.}
\end{figure}

\section{Discussion}
\label{sec-disctextu}

As we see from the figure, the angular power spectrum  
$\ell (\ell + 1) C_{\ell}$  yields the Doppler peaks.
For $\ell <1000$, we find three peaks located
at $\ell=365$, $\ell=720$ and $\ell = 950$. Silk damping, which we
have not taken into account here, will decrease the
relative amplitude of the third peak with respect to the second one;
however it will not affect substantially the height of the first peak.
The integrated SW effect, which also has been neglected, will
shift the position of the first peak to somewhat larger scales, lowering
$\ell_{peak}$ by (5 -- 10)\% and possibly increasing its amplitude slightly  
(by less than 30\%) (Durrer, Gangui \& Sakellariadou 1995).

Our second important result regards the amplitude
of the first Doppler peak, for which we find
$\ell(\ell+1)C_\ell\left|_{{~}_{\!\! \ell\sim 365 }}=5\epsilon^2~.\right.$ 
It is interesting to notice that the position of the first peak is
displaced by  $\Delta \ell\sim 150$  towards smaller
angular scales than in standard inflationary models (Steinhardt 1995).
This is due to the difference in the growth of super--horizon
perturbations (Sugiyama 1996), 
which is $D_r \propto \eta^{3/2}$ in our case,
and $D_r \propto \eta^2$ for inflationary models.

Let us now compare our value for the Doppler peak with the level of the
SW plateau. Unfortunately the numerical value for the SW amplitude
is uncertain within a factor of about 2, which leads to a factor 4
uncertainty in the SW contribution to the power spectrum: 
the results are 
$\ell(\ell+1)C_\ell\left|_{{~}_{\!\! SW}} \sim 2 \epsilon^2
\right.$ (Bennett \& Rhie 1993; Pen et al. 1994)
and
$\ell(\ell+1)C_\ell\left|_{{~}_{\!\! SW}} \sim 8 \epsilon^2
\right.$
(Durrer \& Zhou 1995).
According to the first two groups, the Doppler peak is a factor of $\sim 3.4$
times higher than the SW plateau, whereas it is only about 1.5 times higher if
the second result is assumed. (We allow for about 30\% of the SW
amplitude to be added in phase to the Doppler amplitude of
$\sim 2.24\epsilon$.)
Improved numerical simulations
or analytical approximations are needed to resolve this discrepancy.
However, it is apparent 
that the Doppler contribution from textures is somewhat
smaller than for generic inflationary models.

We believe that our results, about the position and amplitude of  the
first Doppler peak, are basically valid for all global
defects. This depends crucially on the $ 1/\sqrt{\eta}$ behavior of
$(\phi')^2$ on large scales (cf. Eq.~(\ref{fit})).
The displacement of the peaks towards smaller scales 
is reminiscent of open models, where the first Doppler 
peak is located at $\ell_{peak} \simeq 220 / \sqrt{\Omega_0}$.
A value $\Omega_0 \simeq 0.3$ would produce 
$\ell_{peak} \simeq 365$ for the first peak, as we actually find.
 However, the amplitude of the peaks within open inflationary 
models is generically larger than the one we find for global defects.


The problem treated in this paper has also been studied by (Crittenden 
\& Turok 1995).  They obtained the same position of the first Doppler
peak but with a somewhat higher amplitude.

\acknowledgments

One of us (A.G.) is grateful to the organizers of the meeting 
for the invitation to present the seminar.
A.G. thanks the ICTP and the British Council for partial financial support.

\end{document}